\documentclass[12pt,english,preprint,authoryear]{elsarticle}

\usepackage[T1]{fontenc}
\usepackage[latin9]{inputenc}
\usepackage{float}
\usepackage{url}
\usepackage{amsmath}
\usepackage{amssymb}
\usepackage{graphics}
\usepackage{graphicx}
\usepackage{color}
\usepackage{lineno}
\usepackage{babel}

\journal{Icarus}

\begin{document}
\begin{frontmatter}

\title{Origin and Sustainability of The Population of Asteroids Captured
in the Exterior Resonance 1:2 with Mars}

\author[uru]{Tabaré Gallardo\corref{cor1}}
\ead{gallardo@fisica.edu.uy}
\cortext[cor1]{Corresponding author}
\author[uru]{Julia Venturini}
\author[bra]{Fernando Roig}
\author[arg]{Ricardo Gil-Hutton}

\address[uru]{Departamento de Astronomía, Instituto de Física, Facultad de Ciencias,
Iguá 4225, 11400 Montevideo, Uruguay}
\address[bra]{Observatório Nacional, Rua General José Cristino 77, 20921-400,
Rio de Janeiro, Brasil}
\address[arg]{Complejo Astronómico El Leoncito (CASLEO), Av. España 1512 sur,
J5402DSP San Juan, Argentina}

\begin{abstract}
At present, approximately $1500$ asteroids are known to evolve inside
or sticked to the exterior 1:2 resonance with Mars at $a\simeq2.418$
AU, being (142) Polana the largest member of this group. The effect
of the forced secular modes superposed to the resonance gives rise
to a complex dynamical evolution. Chaotic diffusion, collisions, close
encounters with massive asteroids and mainly orbital migration due
to the Yarkovsky effect generate continuous captures to and losses
from the resonance, with a fraction of asteroids remaining captured
over long time scales and generating a concentration
in the semimajor axis distribution
that exceeds by 20\% the population of background asteroids.
The Yarkovsky effect induces different dynamics according to the asteroid
size, producing an excess of small asteroids inside the resonance.
The evolution in the resonance generates a signature on the orbits,
mainly in eccentricity, that depends on the time the asteroid remains
captured inside the resonance and on the magnitude of the Yarkovsky
effect. The greater the asteroids, the larger the time they remain
captured in the resonance, allowing greater diffusion in eccentricity
and inclination. The resonance generates a discontinuity and mixing
in the space of proper elements producing misidentification of dynamical
family members, mainly for Vesta and Nysa-Polana families. The half-life
of resonant asteroids large enough for not being affected by the Yarkovsky
effect is about 1 Gyr. From the point of view of taxonomic classes,
the resonant population does not differ from the background population
and the excess of small asteroids is confirmed.
\end{abstract}

\begin{keyword}
Asteroids, dynamics \sep Asteroid Vesta \sep Resonances, orbital \sep Rotational dynamics
\end{keyword}

\end{frontmatter}


\section{Introduction}

\label{intro}

The asteroid Main Belt is sculpted by mean motion resonances (MMR),
mainly with Jupiter, and secular resonances with the giant planets.
However, there are also relatively strong exterior MMR with the terrestrial
planets, especially in the inner region of the Belt \citep{atlas}.
For any terrestrial planet, its exterior 1:2 MMR is the strongest
one, and in the case of Mars, this resonance (1:2M) is located in
the inner main belt at $a\sim2.4184$ AU. This MMR produces an evident
concentration of asteroids in semimajor axis \citep{gall07}. How
was this concentration created? How long does an asteroid stay in
resonant motion? Is there any mechanism that replenishes the population?
The study of the dynamics of the 1:2M resonance and the origin and
evolutionary paths of their members is the aim of this paper.

If we try to approximate the perturbing function for this resonance
by an analytical approach using a series of terms of different orders
in the small parameters $e$ and $i$, the lowest order terms depend
on the critical angles $\sigma=2\lambda-\lambda_{M}-\varpi$ and $\sigma_{1}=2\lambda-\lambda_{M}-\varpi_{M}$,
where subscript $_{M}$ refers to Mars. These terms are the ones that
are usually librating in the case of asteroids located in the resonance.
The first one is factorized by the asteroid's eccentricity, $e$,
and the second one is factorized by Mars' eccentricity, $e_{M}$.
Higher order terms, and consequently dynamically less important ones,
depend on other combinations of $\lambda,\varpi,\Omega,\lambda_{M},\varpi_{M}$
and $\Omega_{M}$, which define other critical angles. In very particular
cases, these critical angles can librate too.

In this resonance, as in any exterior resonances of the type 1:N,
the librations of the critical angle $\sigma$ occur around central
values, $\sigma_{c}$, that depend on the eccentricity, giving rise
to the so-called {}``asymmetric librations'' because
they do not occur around $0\text{º}$ or $180\text{º}$ as is usual
in other MMRs. The angles $\sigma$ and $\sigma_{1}$, and especially
the first one, are the most important indicators of the resonant state
but not the only ones. In particular, the libration theory predicts
an oscillation of the osculating semimajor axis around a mean, $a_{m}$,
or {}``proper'' value, $a_{p}$, which is characteristic
of the resonance. The difference between $a_{m}$ and $a_{p}$ is
that the first one is the mean over some time span and the latter
is obtained from a frequency analysis. Moreover, it is known that
the resonance width, i.e. the region in semimajor axis where the resonant
motion dominates, is proportional to the orbital eccentricity defining
a region in the $(a,e)$ space where the resonance drives the orbital
evolution.

\begin{figure}[H]
\resizebox{\hsize}{!}{\includegraphics{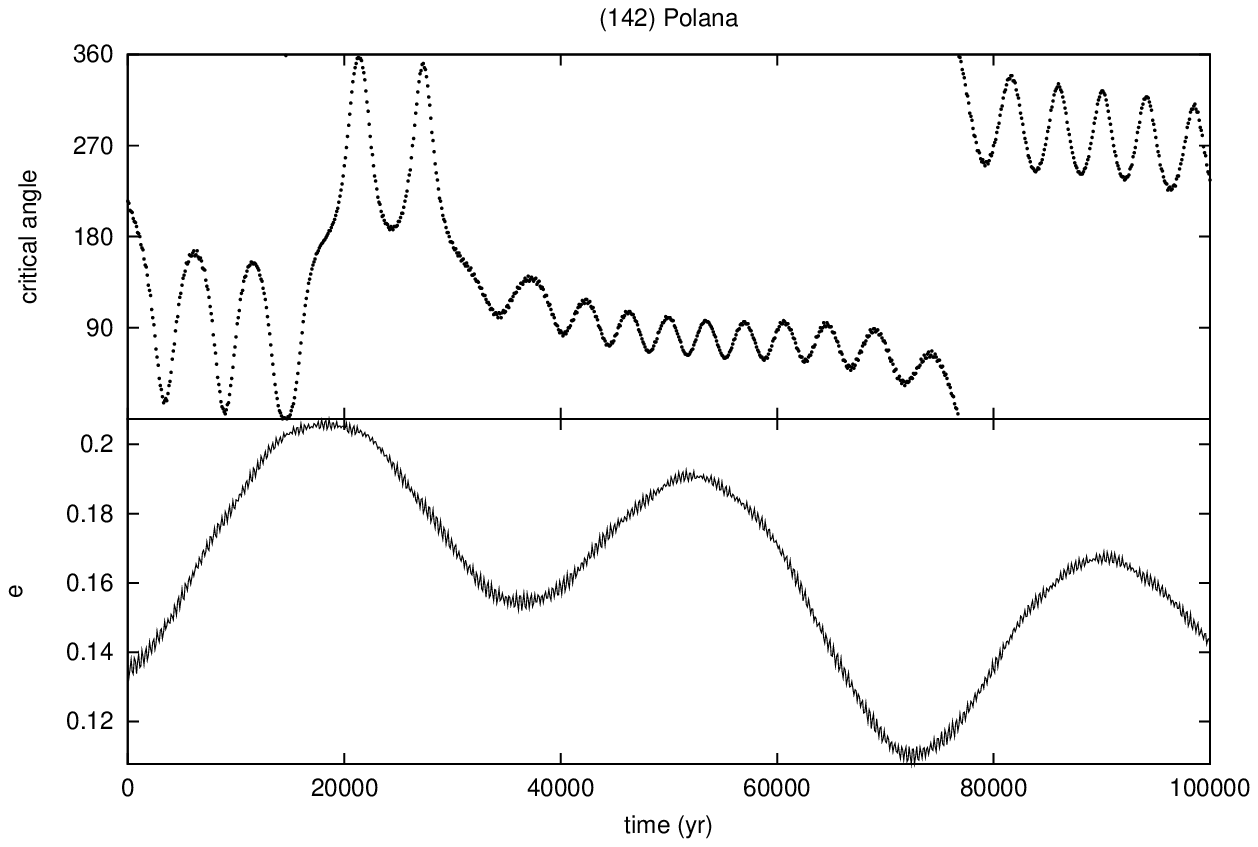}}
\caption{Asymmetric librations of the critical angle $\sigma=2\lambda-\lambda_{M}-\varpi$
of (142) Polana, switching its libration center at $t\backsimeq$
19.000, 30.000 and 77.000 yr. The eccentricity variations are due
to the secular forced modes.}

\label{polana}
\end{figure}

Approximately 1500 asteroids are performing librations, horseshoes
and transitions between both asymmetric libration centers, which are
the typical trajectories in this resonance. About 400 of these asteroids
librate around the asymmetric libration centers with amplitudes
$\Delta\sigma=\sigma_{max}-\sigma_{min}<180^{\circ}$.
The states of the resonant population now and after 1 Myr of orbital
evolution are shown in Figs. 1 and 2 of \citet{gall09}. Asteroids
change their libration center and change between librations and horseshoe
trajectories, but the number of asteroids with asymmetric librations
and horseshoes is similar at the beginning and at the end of this
time-span. Therefore, we may conclude that the population is in equilibrium
over such timescale. Table \ref{biggest} lists the resonant asteroids
with absolute magnitude $H<14$. The evolution of the critical angle
and eccentricity of (142) Polana is shown in Fig. \ref{polana}. Proper
elements shown in the table were derived from our numerical study
which is explained in Section \ref{resonance}, and the complete list
can be found at \url{http://www.fisica.edu.uy/~gallardo/marte12/all.html}.
In some cases, normally at low $e$, the libration amplitude of $\sigma_{1}$
is smaller than the libration amplitude of $\sigma$.

\begin{table}
\centering %
\begin{tabular}{lccccc}
\hline
Asteriod  & $H$  & $e_{p}$  & $i_{p}${[}$^{\circ}${]}  & $\sigma_{c}${[}$^{\circ}${]}  & $\Delta\sigma${[}$^{\circ}${]} \tabularnewline
\hline
(142) Polana  & 10.2  & 0.1573  & 3.198  & 105  & 211 \tabularnewline
(1998) Titius  & 12.2  & 0.0848  & 7.764  & 70  & 124 \tabularnewline
(3665) Fitzgerald  & 12.6  & 0.1151  & 9.557  & 289  & 95 \tabularnewline
(9652) 1996 AF2  & 13.0  & 0.1967  & 6.860  & 96  & 125 \tabularnewline
(2798) Vergilius  & 13.1  & 0.0438  & 5.944  & 214  & 200 \tabularnewline
(11576) 1994 CL  & 13.3  & 0.1394  & 11.111  & 80  & 131 \tabularnewline
(11055) Honduras  & 13.5  & 0.1937  & 11.774  & 70  & 72 \tabularnewline
(11751) 1999 NK37  & 13.7  & 0.1347  & 2.656  & 94  & 142 \tabularnewline
(8748) 1998 FV113  & 13.7  & 0.0917  & 5.937  & 226  & 170 \tabularnewline
(42786) 1998 WU4  & 13.8  & 0.1196  & 22.383  & 141  & 197 \tabularnewline
(2994) Flynn  & 13.9  & 0.1974  & 2.649  & 275  & 102 \tabularnewline
\hline
\end{tabular}\caption{The largest resonant asteroids with critical angles ($\sigma$) showing
asymmetric librations (libration center $\sigma_{c}\neq0\text{º},180\text{º}$)
or transitions between libration centers. $H$ is the absolute magnitude,
$e_{p}$ and $i_{p}$ are proper eccentricity and inclination, and
$\Delta\sigma=\sigma_{\mathrm{max}}-\sigma_{\mathrm{min}}$.}

\label{biggest}
\end{table}

At variance to the resonances with Jupiter that dominate the time
evolution of the asteroid eccentricity, the libration theory predicts
that the eccentricity is almost unaffected by the 1:2M resonance,
but its evolution is dominated by the secular forced modes on the
asteroid orbit generated by the planetary system. In fact, the variations
in eccentricity shown in Fig. \ref{polana} are due to a secular evolution,
not to the libration itself. These important oscillations in the eccentricity
affect the locations of the libration center and the amplitude of
the asymmetric librations of the critical angle (Fig. 2 from \citealp{gall07}).
Nevertheless, these are long term periodic oscillations and they do
not destroy the resonant motion. This can be checked by following
the time evolution of $a_{p}$ or $a_{m}$, which remain locked in
the resonance's domain. However, the irregular evolution of $\sigma$
no longer guarantees the resonant protection mechanism against close
encounters with Mars, allowing such close encounters for high eccentricity
orbits.

The region around the location of the resonance is occupied by the
dynamical families of Vesta, Massalia and the Nysa-Polana complex
\citep{cel01,mot05,zap95}. The 1:2M resonance generates some dynamical
signatures to the Massalia family \citep{vokr06}, and should generate
some dynamical signatures to the other families, a topic that we will
explore in this paper.

The resonance is very narrow in semimajor axis units, but even so
it generates a feature in the distribution of main belt asteroids,
which means that it is strong enough to affect their dynamical evolution.
We are interested in determining the resonance ability to retain asteroids
in spite of the chaotic diffusion, the Yarkovsky effect, and the mutual
collisions and close encounters with massive asteroids that may occur
in that region.

\section{Resonant population: dynamical and physical properties}

\label{resonance}

We have explored the dynamics of the resonance by a series of numerical
integrations using the integrator EVORB \citep{fer02}. Our model
includes the planets from Venus to Neptune, ignoring the planet Mercury,
plus real and fictitious populations of asteroids. The orbital elements
were taken from the ASTORB database (\url{ftp://ftp.lowell.edu/pub/elgb/astorb.html})
and the output of the integrations was analyzed using fourier techniques
\citep{gfm96}. The numerical integrator is basically a leapfrog scheeme
with time step of 0.02 yr in the absence of close encounters, which
switches to a Bulirsch-Stoer routine in the case of close encounters
of the asteroids with the planets. We also analyzed the physical properties
of the asteroids, like absolute magnitude, $H$, surface colors from
the Sloan Digital Sky Survey Moving Objects Catalogue (SDSS-MOC) photometry,
and membership to the known dynamical families in this region.

\subsection{Resonance dynamics}

\begin{figure}[H]
\resizebox{\hsize}{!}{\includegraphics{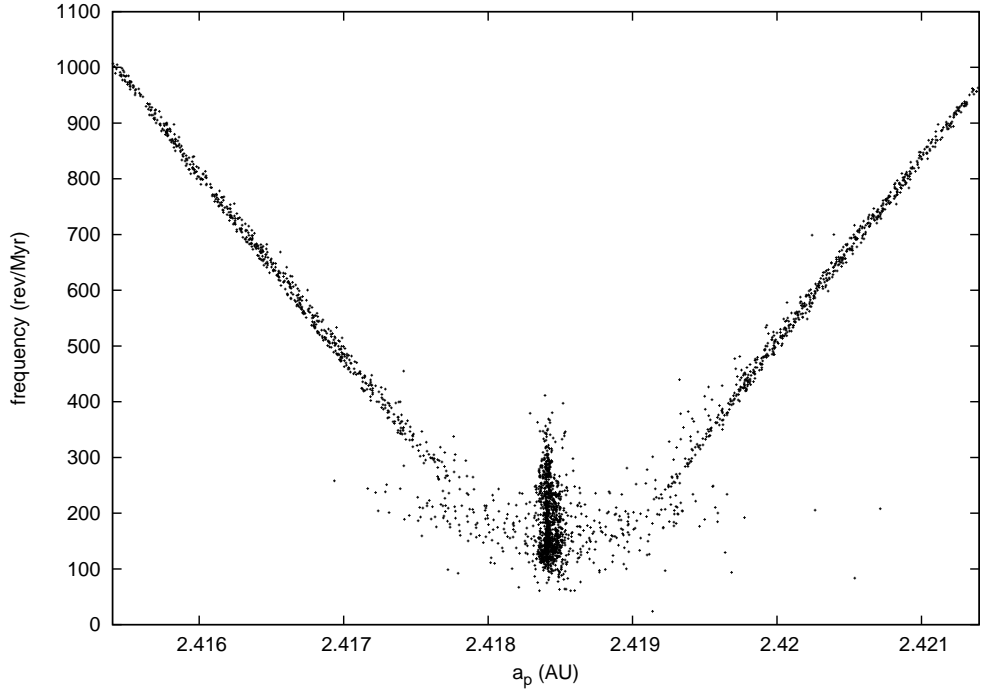}}
\caption{Frequency \textit{vs.} proper semimajor axis at the end of a numerical
simulation of 4022 real asteroids over 30,000 years. Each asteroid
is represented by a dot corresponding to its $a_{p}$ and to the most
representative frequency in the time evolution of $a$. The central
concentration at $a_{p}\simeq2.4184$ corresponds to librations and
horseshoes, the straight lines ($a_{p}\lesssim2.4174$ and $a_{p}\gtrsim2.4194$)
correspond to circulations, and the intermediate zone corresponds
to transitions between horseshoes and circulations.}

\label{aprop-freq}
\end{figure}

In order to find the properties of the resonant motion, we looked
for known asteroids with osculating semimajor axes near the resonant
value and follow their orbital evolution by means of numerical integrations.
We started our analysis by integrating 4022 orbits from the ASTORB
file with $2.416\leq a\leq2.422$ AU. The resonance domain in semimajor
axis can be visualized in several ways. For example, in Fig. \ref{aprop-freq},
we show a plot of $a_{p}$ versus the principal frequency associated
with the time evolution of $a$. Both values were deduced automatically
from the spectral analysis of the time evolution of $a$ and $\sigma$
over $30\times10^{3}$ years. We define $a_{p}$ as the time-independent
term in the fourier series expansion of the osculating $a(t)$. In
Fig. \ref{aprop-freq}, the central feature at $a_{p}\simeq2.4184$
AU corresponds to asteroids performing librations with periods going
from 2,500 to 10,000 yr. The two inclined lines at both sides of the
central feature are due to asteroids with their critical angles performing
circulations. Moving in $a_{p}$ from the left (where $d\sigma/dt>0$)
to the right, the circulation frequency diminishes up to the point
where $d\sigma/dt\sim0$, and the librations start generating the
discontinuity in $a_{p}$ which accumulates at $a_{p}\simeq2.4184$.
Continuing to the right we find, again, the circulation of the critical
angle but in this case with $d\sigma/dt<0$ and with increasing frequency.
Between the central feature and the inclined lines, there are two
regions where transitions between circulation and libration are observed.
For example, an asteroid with $a_{p}=2.416$ shows only circulation
in the critical angle, but this circulation frequency is evident in
the time evolution of $a$, which means that there are traces of the
resonance in its dynamical evolution. No frequencies associated with
$\sigma$ were found for $a_{p}<2.414$ and $a_{p}>2.423$, so no
dynamical traces of the resonance can be expected there.

\begin{figure}[H]
\resizebox{\hsize}{!}{\includegraphics{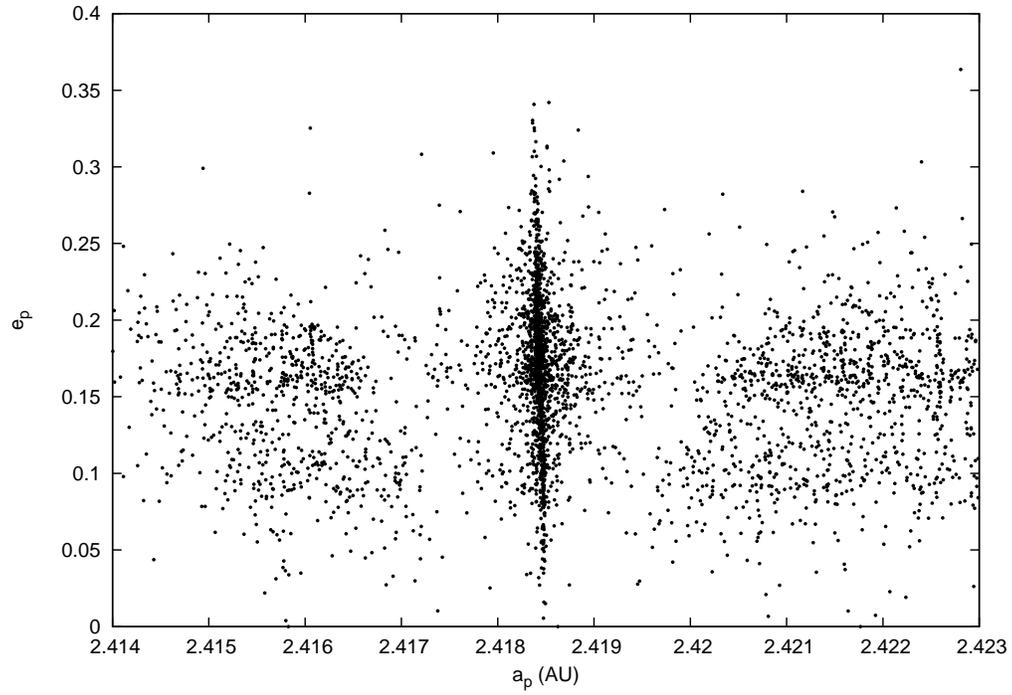}}
\caption{Proper elements obtained from a numerical integration of 4022 real
asteroids over 1 Myr. Each asteroid is represented by a dot corresponding
to its $a_{p}$ and $e_{p}$. The central region corresponds to librations
and horseshoes. The increasing width of the resonance as a function
of $e_{p}$ is evident.}

\label{aeprop}
\end{figure}

Performing a spectral analysis of the variables $(k=e\cos\varpi,\, h=e\sin\varpi)$
and $(p=i\cos\Omega,\, q=i\sin\Omega)$ over 1 Myr of numerical integration,
it was possible to determine the forced and proper modes in the evolution
of $e$ and $i$. Forced modes are the oscillation modes corresponding
to the fundamental frequencies of the planetary system, and proper
modes are the oscillation modes with large amplitudes and frequencies
different from the fundamental ones \citep{mude}. We call $e_{p}$
and $i_{p}$ the amplitudes of the forced modes. Contrary to Fig.
\ref{aprop-freq}, where we cannot distinguish between different eccentricities,
Fig. \ref{aeprop} shows the structure of the resonance with increasing
width at higher $e_{p}$, and the huge concentration of asteroids
at $2.4180<a_{p}<2.4188$, with two deep gaps at both sides of the
central peak. An analogous plot using synthetic proper elements from
AstDyS (\url{http://hamilton.dm.unipi.it/astdys/}) looks very similar
to our Fig. \ref{aeprop}. Considering the peak and the gaps, there
is an excess of asteroids in the region $2.416<a_{p}<2.421$. According
to the AstDyS database, that only includes asteroids with well determined
orbits so it should be less biased than our sample, we estimate that
in a range of $\Delta a_{p}\simeq0.005$ AU there are approximately
1900 asteroids in the domain of the resonance, against approximately
1600 asteroids that are far from the resonance (background). Therefore,
in that database, there is an excess of approximately 300 resonant
asteroids, which means that the resonance generates a population excess
of 20\% over the background.

\begin{figure}[H]
\resizebox{\hsize}{!}{\includegraphics{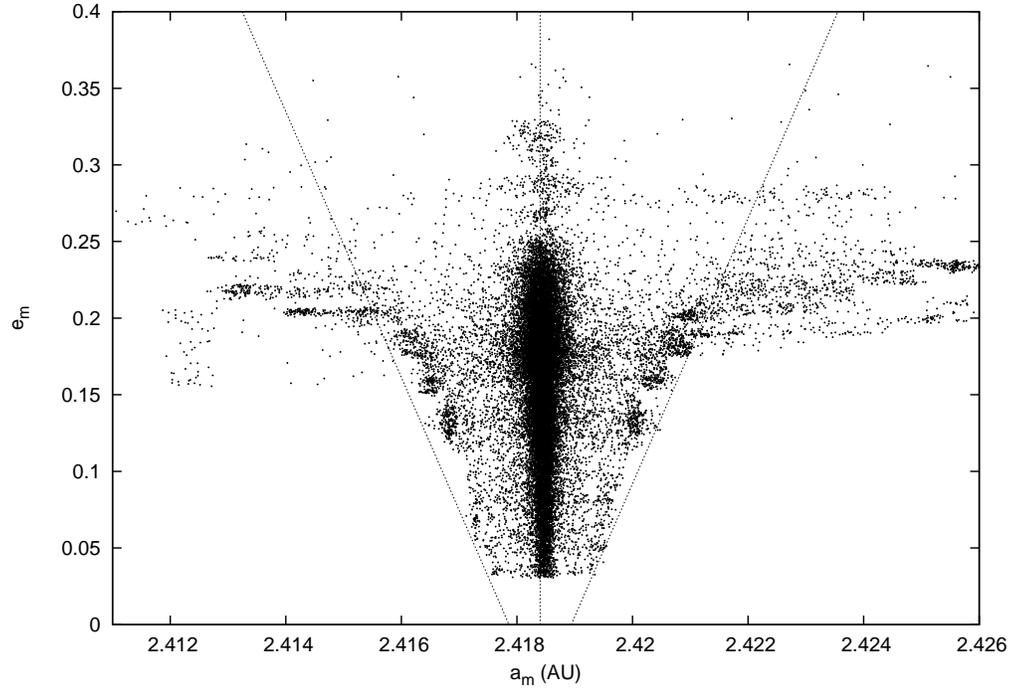}}
\caption{Mean orbital elements, obtained from a 1 Myr running window, of 100
resonant asteroids from the ASTORB database numerically integrated
over 1 Gyr. The concentration at $a_{m}\simeq2.4184$ AU is due to
librations and horseshoes. The concentration at the borders of the
resonance is due to the phenomenon of stickiness. The behavior at
$e_{m}>0.2$ is due to chaotic diffusion dirven by the overlap of
several high order resonances. The limits of the resonance are approximately
given by the shaded lines. (Low resolution version)}

\label{limites100}
\end{figure}

Figures \ref{aprop-freq} and \ref{aeprop} can be used to define
the limits where the resonance is dominant over the other periodic
terms that affect the time evolution of the asteroid semimajor axis.
However, the best indicator of these limits appears in Fig. \ref{limites100},
that was obtained by plotting the 1 Myr averages of the orbital elements
of 100 resonant asteroids from the ASTORB file evolving over 1 Gyr.
The concentration at the borders of the resonance is due to the phenomenon
of {}``resonance stickiness'' \citep{matr99}, and the behavior
at $e_{m}>0.2$ is due to the overlap of resonances that cause a strong
chaotic diffusion \citep{mn99}.

\begin{figure}[H]
\resizebox{\hsize}{!}{\includegraphics{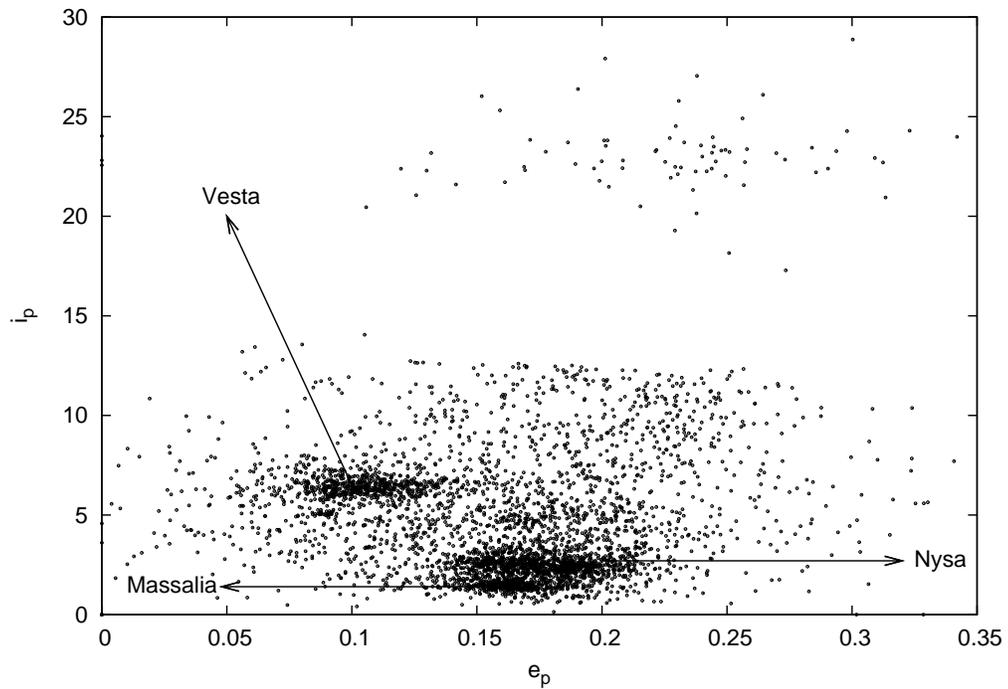}}
\caption{Proper elements of the population of 4022 asteroids inside and close
to the resonance. It is possible to identify three clusters that can
be associated to the families of Vesta, Nysa-Polana and Massalia.
See Table \ref{tabla0} for details. The empty region over $i_{p}\sim15^{\circ}$
is due to the secular resonance $\nu_{6}$.}

\label{eiprop}
\end{figure}

The fundamental frequencies $g_{5}$ and $g_{6}$ are clearly present
in the secular evolution of the eccentricity, and $f_{6}$ appears
in the inclination. The proper elements $e_{p}$ and $i_{p}$ are
concentrated around some values, indicating their relation to particular
collisional families (Fig. \ref{eiprop}).

\subsection{Physical properties}

\label{phypro}

According to the orbital evolution over 1 Myr, we defined a subsample
of asteroids clearly captured in the resonance ($2.4180<a_{p}<2.4188$)
and a subsample of asteroids clearly outside of both sides of the
resonance ($a_{p}<2.4154$ and $a_{p}>2.4214$). In order to determine
the fraction of these bodies that belong to asteroid dynamical families,
we applied the Hierarchical Clustering Method (following the procedure
described in \citealp{mot05}) to the AstDyS catalog of analytical
proper elements, and identified the major asteroid families in the
neighborhood of the 1:2M resonance. We found that only 11.5\% of the
clearly resonant asteroids belong to the dynamical families of either
Vesta, Nysa-Polana or Massalia. For the non resonant subsample, only
18.9\% are members of any of the above families (see details in Table
\ref{tabla0}). The families of Vesta and Nysa-Polana clearly have
less members inside the resonance than outside. Results for Nysa-Polana
family are consistent with Fig. 26 of \citet{mal95}, where a decrease
in the density number of the family members is observed at $a\simeq2.42$.
As we will show later, the lack of family members in the resonance
is related to the diffusion in $(e,i)$ that hinders the identification
of family members \citep{vokr06,cami08}.

\begin{table}
\centering

\begin{tabular}{lccccc}
\hline
Family  & $a_{p}${[}AU{]}  & $e_{p}$  & $i_{p}${[}$^{\circ}${]}  & \% of R  & \% of NR \tabularnewline
\hline
Vesta  & 2.25 - 2.48  & 0.08 - 0.12  & 5.7 - 7.5  & 2.9  & 6.2 \tabularnewline
Nysa-Polana  & 2.32 - 2.47  & 0.14 - 0.20  & 2.0 - 3.2  & 5.3  & 9 \tabularnewline
Massalia  & 2.36 - 2.46  & 0.15 - 0.17  & 1.3 - 1.7  & 3.3  & 3.7 \tabularnewline
\hline
\end{tabular}

\caption{Percentage of dynamical family members in the subsample of resonant
asteroids (R) and in the subsample of non resonant asteroids (NR).
There is a deficiency of identified family members inside the resonance.}

\label{tabla0}
\end{table}

\begin{figure}[H]
\resizebox{\hsize}{!}{\includegraphics{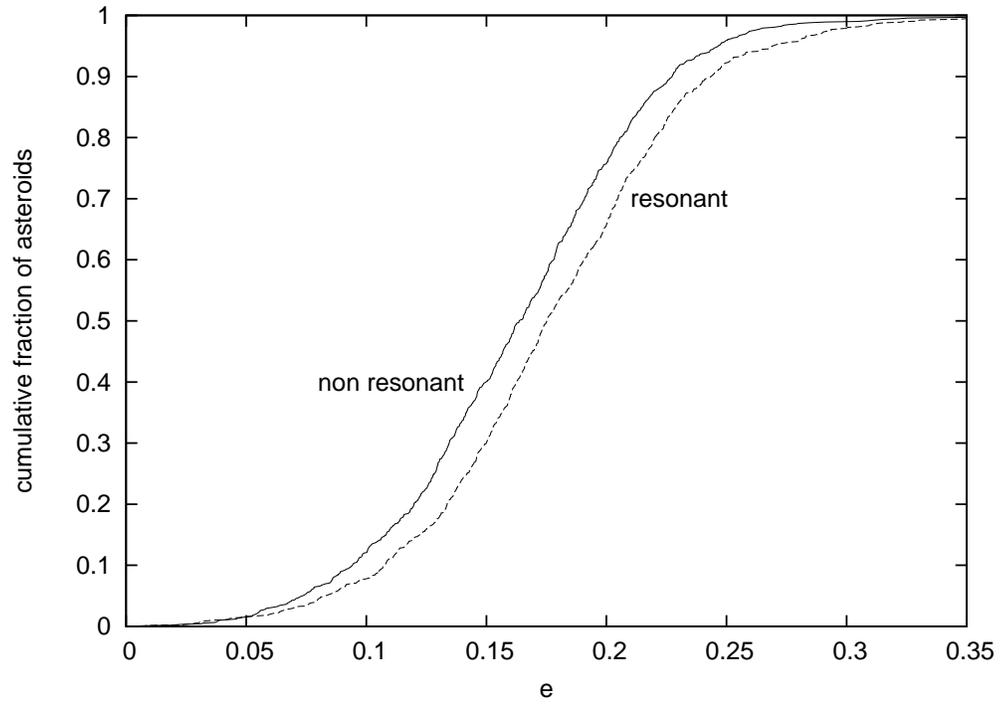}}
\caption{Cumulative distribution of the eccentricities for the resonant and
non resonant asteroids. The resonant asteroids have comparatively
more eccentric orbits than the non resonant ones.}

\label{histexc}
\end{figure}

\begin{figure}[H]
\resizebox{\hsize}{!}{\includegraphics{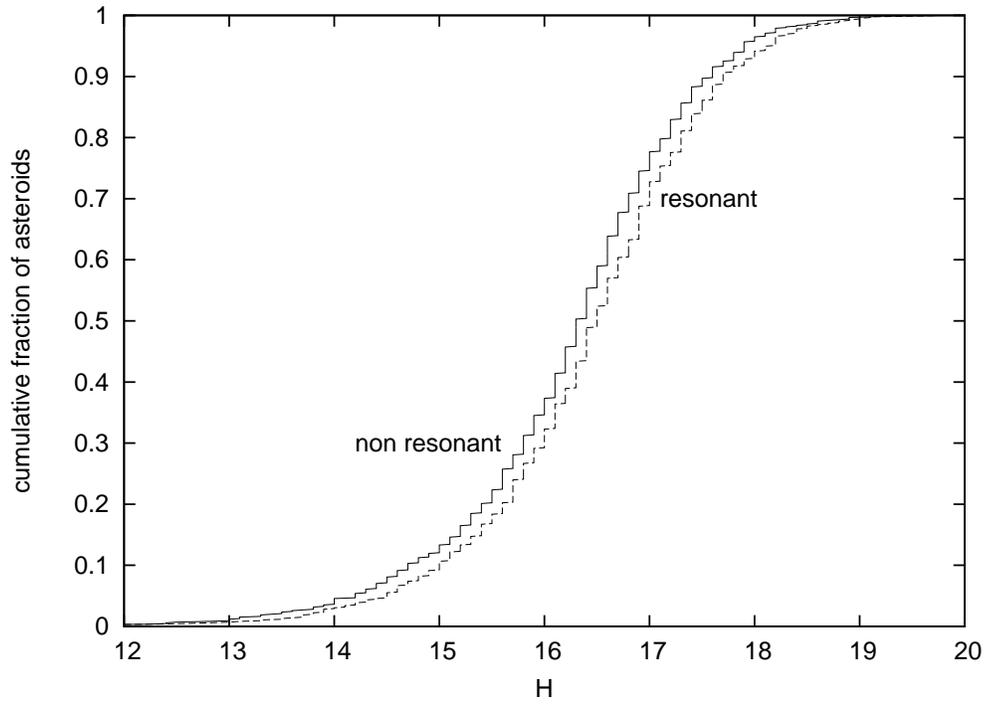}}
\caption{Cumulative distribution of the absolute magnitudes for two unbiased
samples of resonant and non resonant asteroids. The two samples of
855 asteroids each, have the same eccentricity distribution in order
to avoid possible discovery biases due to different perihelion distances.
The resonant asteroids are comparatively smaller than the non resonant
ones.}

\label{H}
\end{figure}

For both resonant and non resonant groups, we analyzed the distribution
of asteroids' surface colors using the photometric data from Sloan
Digital Sky Survey (SDSS). We did not find any remarkable differences between
the groups, which led us to conclude that the resonance is not dominated
by any particular asteroid taxonomic class. The analysis of the absolute
magnitude ($H$) distribution confirmed a tendency to higher $H$
(smaller asteroids) for the resonant population \citep{gall07}. However,
the resonant population is more eccentric, on average, than the non
resonant one (Fig. \ref{histexc}), and this might favor the discovery
of smaller asteroids inside the resonance than outside. To correct
this bias, we analyzed the $H$ distribution of two samples of 855
asteroids each, one from the resonant population and the other from
the non-resonant population, having the same eccentricity distribution
in the range $0<e<0.25$ (Fig. \ref{H}). Using a Kolmogorov-Smirnov
test for cumulative distributions \citep{nure}, we found that both
samples have different $H$ distributions with a level of confidence
of $98.7\%$, and that small asteroids are more abundant in the resonant
sample. We will discuss this point later in Section \ref{secyar}.

\subsection{Long term evolution of the resonant population}

\begin{figure}[H]
\resizebox{\hsize}{!}{\includegraphics{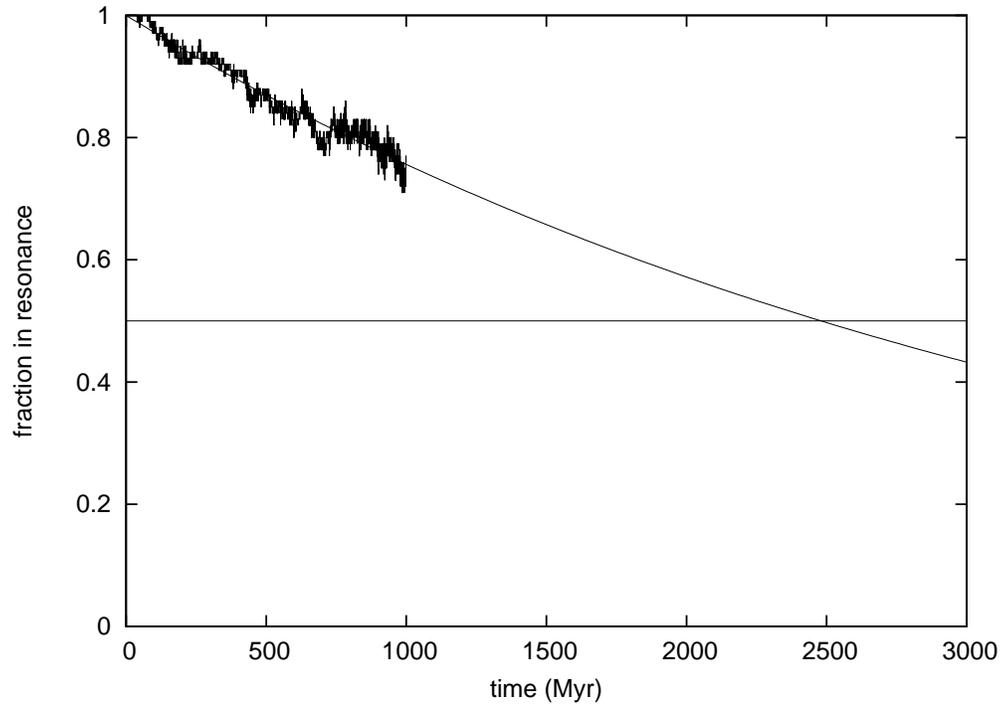}}
\caption{Time decay of the sample of 100 asteroids shown in Fig. \ref{limites100},
initially in resonant motion. A power law fit gives a half-life inside
the resonance of 2.5 Gyr.}

\label{vidamedianoy2}
\end{figure}

We integrated numerically 100 real asteroids over 1 Gyr, and calculated
the mean eccentricity and the mean semi-major axis over intervals
of 1 Myr. Then, using the limits of the 1:2M resonance defined by
Fig. \ref{limites100}, we determined the number of asteroids captured
in the resonance at each instant and the result is shown in Fig. \ref{vidamedianoy2}.
By fitting this data to an exponentially decaying function, we can
deduce a half-life of $\sim2.5$ Gyr. This value will be revised later
in Section \ref{secyar}.

While evolving inside the resonance, the asteroids exhibit a diffusion
mainly in $e_{m}$, but also in $i_{m}$, which in general is roughly
proportional to the lifetime in the resonance. A typical variation
in $e_{m}$ is about $0.05$ after some hundred million years of evolution
inside the resonance, but some asteroids may exhibit larger changes
in $e_{m}$, and variations up to 0.15 can be achieved for the asteroids
that remain captured over 1 Gyr. We have also found that the larger
proper inclination, the larger the diffusion in $e_{m}$ and $i_{m}$.
For example, for $i_{p}=1.5^{\circ}$, typical of the Massalia family,
the diffusion in eccentricity is 0.05 and in inclination is $1.2^{\circ}$,
whereas for $i_{p}=6.6^{\circ}$, typical of the Vesta family, the
diffusion in eccentricity is 0.08 and in inclination is $1.6^{\circ}$.
The success in the identification of resonant members of the Massalia
family, in comparison to the members of the Nysa-Polana family, could
be explained partially because of their low proper inclination and,
in consequence, their lower diffusion in eccentricity and inclination.
Figure \ref{einoy} shows the evolution in $(e_{m},i_{m})$ for the
complete sample. When $e_{m}$ crosses the value $\sim0.22$, the
diffusion in $a_{m}$ and $i_{m}$ increases and the asteroids evolve
in a chaotic region generated by the overlap of several resonances,
eventually leaving the 1:2M resonance.

In our simulation, the first asteroid had a close encounter with Mars
after 84 Myr, and after 200 Myr there were 5 asteroids encountering
Mars, so the resonance is a clear source of Mars crossing asteroids.
The first encounter of an asteroid with the Earth happened at approximately
100 Myr. Six asteroids in the simulation, most of them with initial
$i\gtrsim11^{\circ}$, ended by colliding with the Sun shortly after
being captured into the 3:1 resonance with Jupiter (the most common
situation), or in the 4:1 resonance with Jupiter, or evolving near
the $\nu_{6}$ secular resonance.

Other MMRs very close to the 1:2M resonance are listed in Table \ref{tabla2bmmr}.
These resonances have much lower strengths, however, for higher values
of $e$ they grow in relevance and can compete with the 1:2M resonance,
allowing a diffusion in $a$. We have also found captures in three
body mean motion resonances \citep{nemo98}, as for example the $3n_{M}+n_{N}-6n=0$
resonance at $a\simeq2.4125$, that involves the mean motions of Mars
and Neptune. All these MMRs contribute to the chaotic diffusion observed
at high eccentricities.

\begin{table}
\centering

\begin{tabular}{cccc}
\hline
Resonance  & $a${[}AU{]}  & strength $(e=0.15)$  & strength $(e=0.25)$ \tabularnewline
\hline
$19\lambda_{J}-6\lambda$  & 2.4116  & $4\times10^{-7}$  & $1\times10^{-4}$ \tabularnewline
$-4\lambda_{E}+15\lambda$  & 2.4137  & $6\times10^{-6}$  & $5\times10^{-3}$ \tabularnewline
$-\lambda_{M}+2\lambda$  & 2.4184  & 1  & 1 \tabularnewline
$22\lambda_{J}-7\lambda$  & 2.4237  & $2\times10^{-8}$  & $1\times10^{-5}$ \tabularnewline
$-5\lambda_{E}+19\lambda$  & 2.4351  & $1\times10^{-6}$  & $6\times10^{-4}$ \tabularnewline
\hline
\end{tabular}

\caption{The strongest two body mean motion resonances near $a\sim2.418$.
The resonance strength was calculated following \citet{atlas} and
is given in units relative to the 1:2M resonance strength for two
different values of the orbital eccentricity and for fixed $i=5^{\circ}$
and $\omega=60^{\circ}$. Note the increasing strength of the resonances
relative to 1:2M for the more eccentric orbits.}

\label{tabla2bmmr}
\end{table}

\section{Orbital migration by the Yarkovsky effect}

\label{secyar}

All the results presented up to this point were obtained in the framework
of pure gravitational interaction between the asteroids and the planets.
It is known \citep{bot2000} that the Yarkovsky effect generates a
small extra acceleration due to the natural delay of thermal emission
by the asteroid surface, and this effect could significantly modify
the capture mechanisms, especially for asteroids with diameter $\lesssim10$
km.

We found that 80\% of the 4022 asteroids located in the 1:2M resonance
region have absolute magnitudes $14.9<H<17.8$, which corresponds
to diameters between $\sim0.6-6.0$ km depending on the geometrical
albedo considered. So, we are dealing with a population of asteroids
for which the Yarkovsky effect cannot be neglected. This effect can
be reproduced with a simplified model by adding a term to the asteroid's
acceleration of the form \citep{roig07}:
\begin{equation}
\ddot{\vec{r}}=\frac{GM_{\odot}\dot{a}}{2a^{2}}\frac{\vec{v}}{v^{2}}\label{model}
\end{equation}
where $G$ is the gravitational constant, $M_{\odot}$ the solar mass,
$a$ the osculating semimajor axis of the asteroid, $\vec{r}$ and
$\vec{v}$ are its heliocentric position and velocity, respectively,
and $\dot{a}$ the assumed Yarkovsky drift in $a$ that depends on
several factors like diameter, rotational axis obliquity, rotational
period, and surface thermal conductivity, $K$. According to Bottke
et al. \citeyearpar{bot2000,bot2006}, $\dot{a}$ runs from $\sim10^{-2}$
to $\sim10^{-5}$ AU/Myr for asteroids with radius from some meters
to a few kilometers and values of $K$ between 0.01 and 0.001. The
most common variant is the Yarkovsky {}``diurnal'' effect, that
makes asteroids with prograde rotation to have $\dot{a}>0$, and those
with retrograde rotation to have $\dot{a}<0$. In very particular
cases, like asteroids with iron-rich surfaces and high values of the
thermal conductivity, another variant known as the Yarkovsky {}``seasonal''
effect (that we will not consider here) predominates and produces
a secular negative drift in $a$ \citep{bot2006}.

We have performed a series of four simulations
with four different assumed values
of $\dot{a}$: $10^{-2}$, $10^{-3}$,
$10^{-4}$ and $10^{-5}$ AU/Myr (hereafter referred as simulations
\#1, \#2, \#3 and \#4, respectively, see Table \ref{tabla1}). In
each simulation, we consider two fictitious population of 200 test
particles initially located at both sides of the resonance and with
values of $(e,i)$ taken from the real population. In order to force
the particles to fall into the resonance, we assumed $\dot{a}>0$
(i.e, prograde rotation) for the particles starting from {}``the
left side'' of the resonance, and $\dot{a}<0$ (i.e., retrograde
rotation) for the particles starting from {}``the right side''.
This allows us to detect any difference between left and right side
populations regarding the capture into the resonance. In these simulations,
we do not consider the abrupt and stochastic changes in $\dot{a}$
due to eventual reorientation of spin axes related to non catastrophic
collisions. We also ignore the obliquity and rotation rate evolution
due to the YORP effect \citep{voca02}. It must be noted that the
simplified model used to simulate the Yarkovsky effect generates a
spurious variation in the particles eccentricity, but this is negligible
in magnitude. Indeed, if we calculate the total variation in $e$
induced by the Eq, (\ref{model}) during our simulations only by means
of the Gauss planetary equations, we obtain $\sim10^{-4}-10^{-3}$
which is small enough to not alter our conclusions.

\begin{table}
\centering %
\begin{tabular}{ccccccc}
\hline
Simulation  & $\dot{a}$ {[}AU/Myr{]} & $R$ {[}m{]}  & $\Delta T$ {[}Myr{]}  & $\Delta a$ {[}AU{]}  & $t_{\mathrm{ejec}}$ {[}Myr{]}  & \tabularnewline
\hline
\#1  & 0.01  & 1  & 2  & 0.02  & 5 C  & \tabularnewline
\#2  & 0.001  & 30  & 20  & 0.02  & 10 C  & \tabularnewline
\#3  & 0.0001  & 600  & 200  & 0.02  & 90 C  & \tabularnewline
\#4  & 0.00001  & 5000  & 1000  & 0.01  & 1000 E  & \tabularnewline
\hline
\end{tabular}\caption{Parameters of four simulations using different Yarkovsky drifts $\dot{a}$.
Each simulation was composed of two fictitious populations of 200
bodies. Each population was initially located at each side of the
resonance. The typical radius of the bodies, $R$, was estimated according
to \citet{bot2006} and assuming a thermal conductivity between 0.01
and 0.001. Table gives the total time span, $\Delta T$, of the simulations
and the equivalent total drift, $\Delta a$. The ejection timescale,
$t_{\mathrm{ejec}}$, has a large uncertainty and it was estimated
as the mean time necessary for the particles to be ejected from the
resonance due to either gravitational perturbations by close encounters
with massive asteroids (labeled as {}``E''), or collisions with
other asteroids, including disruption events (labeled as {}``C'').}

\label{tabla1}
\end{table}

The simulations spanned a time scale long enough for the total drift
in $a$ to be larger than the resonance width (i.e., the asteroids
would completely cross the resonance during the simulations provided
that no captures occur). We also estimated the time scale of the collisional/gravitational
lifetime, $t_{\mathrm{ejec}}$, as the time necessary to eject an
asteroid from the resonance due to collisions with other asteroids,
or due to gravitational perturbations by close encounter with massive
asteroids. To compute $t_{\mathrm{ejec}}$, we simulated collisions
using a Monte Carlo method. We assumed a typical power law population
of projectiles \citep{fa98} with randomly oriented encounter velocities,
$v_{e}$, taken from a Maxwellian distribution with mean $5.3$ km/s.
We also assumed an impulse $\delta v=(R_{\mathrm{proj}}/R_{\mathrm{target}})^{3}v_{e}$,
and computed $\delta a$ via Gauss planetary equations, considering
disruption events as in \citet{bot2005}.

Gravitational perturbations by massive asteroids \citep{neal02} were
also estimated via Gauss planetary equations, using a Monte Carlo
method to calculate the impulse $\delta v=2mGv_{e}(G^{2}(M+m)^{2}+\rho^{2}v_{e}^{4})^{-1/2}$
due to hyperbolic encounters with the 10 most massive asteroids, where
$\rho$ is the randomly chosen impact parameter, $M$ the mass of
the target assuming a density of 2.7 g/cc, and $m$ the mass of any
of the randomly chosen massive asteroids. The mean time interval between
encounters with the 10 most massive asteroids and with impact parameter
$\rho<0.01$ AU was taken to be 15,600 yr. This time scale was deduced
from the intrinsic collision probability of $2.86\times10^{-18}$
km$^{-2}$yr$^{-1}$, typical for the inner region of the Main Belt.

The gravitational perturbations by massive asteroids are approximately
the same for all the range of asteroid sizes. On the other hand, the
effect of collisions is only relevant for smaller asteroids. In our
simulations, we found that $t_{\mathrm{ejec}}$ is determined by collisions
for asteroids with $R\lesssim2.5$ km and by close encounters with
the massive asteroids otherwise. These close encounters drop off the
half-life obtained from Fig. \ref{vidamedianoy2} to $\sim1$ Gyr.

\subsection{Capture in resonance and chaotic diffusion}

The analysis of the time evolution of the semimajor axis shows that
the main mechanism that keeps the asteroid fixed in $a$ is the resonance
stickiness. Figure \ref{p1119} shows a typical example of a particle
from the {}``left side'' group in simulation \#4. The Yarkovsky
effect generates a positive drift in semimajor axis up to $t\simeq500$
Myr, when the particle gets sticked to the resonance alternating between
the {}``left'' and {}``right'' sides. At $t\simeq750$ Myr, the
particle gets captured into the resonance, and after that it evolves
switching between sticking and getting inside the resonance. It is
worth noting that after $t\simeq500$ Myr the drift in $a$ stopped,
and a new dynamical mechanism drives the eccentricity. This mechanism
partially erases the {}``memory'' of the particle, mainly its proper
eccentricity but also its proper inclination. Captures from the {}``right
side'' exhibit a similar behavior. In some cases, the particle is
neither captured into the resonance nor in a sticky orbit, and in
these cases the eccentricity is not altered. This is illustrated from
the results of simulation \#4 shown in Figs. \ref{1gright} and \ref{1gleft2}.
Figure \ref{1gright} shows the population evolving from the {}``right
side'' in the $(a_{m},e_{m})$ plane, and Fig. \ref{1gleft2} shows
the population evolving from the {}``left side'' in the $(e_{m},i_{m})$
plane; the diffusion in this later case is significant for asteroids
captured into resonance and negligible for asteroids that quickly
escape from the resonance.

Calculating the $a_{m}$ of each particle in a 1 Myr running window,
we computed the number of asteroids inside the resonance or sticked
to the resonance as a function of time. We also computed the time
$t_{\mathrm{res}}$ that each asteroid remains captured into/sticked
to the resonance. We considered the asteroid was captured/sticked
when $a_{m}$ falls inside the resonance region in the $(a_{m},e_{m})$
space defined in Fig. \ref{limites100}. Analyzing all four simulations,
we found that the total effect on the eccentricity when evolving inside
the resonance is somehow proportional to the magnitude of the Yarkovsky
effect and to the time the particle remained captured into the resonance
(Fig. \ref{sim4der}). The diffusion in mean eccentricity, $\Delta e_{m}$,
as a function of the lifetime in the resonance $t_{\mathrm{res}}$,
in Myr, can be approximately described by
\begin{equation}
\Delta e_{m}\sim6\hspace{1mm}\dot{a}\hspace{1mm}t_{\mathrm{res}}^{\beta}\label{difusion}
\end{equation}
 where $\dot{a}$ is in AU/Myr and $\beta$ is 0.54, 0.75 and 1.07
for simulations \#2, \#3 and \#4, respectively. The value of $\Delta e_{m}$
deduced from this expression is only an indication of the actual value,
considering the large spread of the data observed in Fig. \ref{sim4der}.
From this figure, we can conclude that, in general, the largest diffusion
in eccentricity corresponds to simulation \#4, i.e., to the largest
asteroids.

Simplified analytical models (see for example \citealp{mude}) predict
that captures from {}``approaching'' orbits (i.e., asteroids migrating
towards Mars, or equivalently, from the {}``right side'' population)
are more successful than captures from {}``receding'' orbits (i.e.,
from the {}``left side'' population), but in our simulations we
did not find any significant difference between the two populations
regarding the probability of capture. As an example, Fig. \ref{evo200}
shows the fraction of asteroids captured into resonance as a function
of the time in simulation \#3 (i.e., for a Yarkovsky effect corresponding
to an asteroid size of some hundred meters), distinguishing between
the left and right populations. The same behavior was obtained in
all the four simulations, but in different time scales.

Using the values of $t_{\mathrm{res}}$ computed for all particles
in all the four simulations, we determined the cumulative distribution
of the \textit{scaled} lifetime in the resonance, $L=t_{\mathrm{res}}/T$,
where $T$ is the time that would be necessary to cross the resonance
region due to the Yarkovsky drift in $a$ only (i.e., assuming that
no captures occur). The result is shown in Fig. \ref{4vidas}. If
$L<1$ in all the simulations, the distribution of the asteroid semimajor
axes would show a gap or a depression at the resonance location. On
the other hand, asteroids with $L>1$ would be the ones contributing
to the total excess of asteroids inside the resonance. Looking at
the left part of Fig. \ref{4vidas} ($L<1$), it is evident that the
smaller the asteroids the larger the scaled lifetime, therefore, an
excess of small asteroids inside the resonance with respect to the
background should be expected. This is consistent with our analysis
of the size distributions (Section \ref{phypro} and Fig. \ref{H}).
It is worth mentioning that \citet{tsig03} also found an excess of
small asteroids in the 7:3 resonance with Jupiter due to a selective
dynamical process generated by the Yarkovsky effect.

The higher capture probability at larger eccentricities, and the diffusive
process in eccentricity inside the resonance, explain the different
eccentricity distributions observed in the resonant population compared
to the non resonant background population.

\subsection{V-type asteroids as tracers}

V-type asteroids have unique surface colors, and they are considered
to be members of a dynamical family formed as a result of an impact
on Vesta's surface. There are several V-type asteroids far from the
domain of the Vesta dynamical family in the space of proper elements,
a fact that has challenged the understanding of the dynamical link
between these asteroids and the Vesta family \citep{neal08}.

\citet{rogh06}, at their Table 2, provide a list of candidate V-type
asteroids that were not identified as members of the Vesta dynamical
family. From that list, we identified seven asteroids evolving in
the 1:2M resonance, whose dynamical parameters are shown in Table
\ref{tablacandidatos}. These asteroids are very close to the region
of the Vesta dynamical family, so they could have evolved in $e_{p}$
due to the diffusive process inside the resonance, masking their dynamical
origin as Vesta family members. The only exceptions are probably (3331)
Kvistaberg and (109721) 2001 RE54, because their $i_{p}$ could not
be explained by a diffusive process inside the resonance according
to our numerical studies (see for example Fig. \ref{einoy}).

\begin{table}
\centering

\begin{tabular}{lcccccc}
\hline
Asteroid  & $a_{p}${[}AU{]}  & $e_{p}$  & $i_{p}${[}$^{\circ}${]} & $H$  & $\sigma_{c}${[}$^{\circ}${]}  & $\Delta\sigma${[}$^{\circ}${]} \tabularnewline
\hline
(3331) Kvistaberg  & 2.41875  & 0.114  & 3.58  & 13.2  & 160  & 360 \tabularnewline
(5599) 1991 SG1  & 2.41869  & 0.155  & 7.23  & 12.5  & 175  & 360 \tabularnewline
(12027) 1997 AB5  & 2.41840  & 0.174  & 6.22  & 14.6  & 183  & 287 \tabularnewline
(40733) 1999 SM17  & 2.41846  & 0.113  & 6.80  & 15.0  & 167  & 340 \tabularnewline
(68064) 2000 YU67  & 2.41893  & 0.152  & 7.13  & 15.2  & 179  & 360 \tabularnewline
(122441) 2000 QH126  & 2.41837  & 0.100  & 6.92  & 15.4  & 177  & 279 \tabularnewline
(109721) 2001 RE54  & 2.41848  & 0.051  & 3.93  & 16.0  & 217  & 180 \tabularnewline
(2798) Vergilius  & 2.41848  & 0.044  & 5.94  & 13.1  & 214  & 200 \tabularnewline
(32366) 2000 QA142  & 2.41840  & 0.213  & 6.31  & 13.9  & 158  & 292 \tabularnewline
\hline
\end{tabular}\caption{Resonant V-type asteroids not identified as members of Vesta dynamical
family. The first seven ones are from \citet{rogh06}. $a_{p}$, $e_{p}$
and $i_{p}$ are the proper semimajor axes, eccentricities and inclinations,
respectibely, $H$ is the absolute magnitude, and $\sigma_{c}$ and
$\Delta\sigma$ are the libration center and libration amplitude,
respectively.}

\label{tablacandidatos}
\end{table}

Looking at the photometric data from the SDSS, we identified other
two candidate V-type asteroids among the resonant population: (32366)
2000 QA142 and (2798) Vergilius. The orbital evolution of these two
asteroids, together with (12027) 1997 AB5, are shown in Fig. \ref{vesto}.
The low eccentricity of (2798) and the high eccentricity of (12027)
are both compatible with the diffusive processes inside the resonance.

The case of (32366) is a very special one, because its $e_{p}$ is
larger by approximately $0.1$ with respect to the value of $e_{p}$
expected for the V-type asteroids. Assuming a geometric albedo of
$\sim0.3-0.4$, typical of V-types, the diameter of asteroid (32366)
is $\sim3.5-4.0$ km. This size is consistent with a drift rate $|\dot{a}|\approx1.25-1.43\,\times10^{-5}$
AU/Myr \citep{roig07}, so we can apply Eq. (\ref{difusion}) or look
at Fig. \ref{sim4der} to conclude that the necessary timescale of
resonant evolution to increase the eccentricity by 0.1 is approximately
800 Myr. In fact, Fig. \ref{vesto} shows that this asteroid has important
excursions in $(e_{m},i_{m})$; thus, it could have obtained its present
eccentricity after several hundred millon years of resonant evolution.

\section{Discussion and Conclusions}

Our analysis of the size distribution of the real asteroids in the
region of the 1:2M resonance has confirmed a significant lack of low
$H$ asteroids in the resonant population compared to the non resonant
one. This result is consistent with the ratio of small to big asteroids
inside and outside the resonance that we obtained from our simulations
of fictitious asteroids, using different values of the drift in $a$
induced by the Yarkovsky effect according to different asteroid sizes.
It is worth mentioning that pure gravitational processes, like chaotic
diffusion, do not distinguish between asteroid sizes; therefore in
the absence of the Yarkovsky effect the size distribution in the resonant
population should be the same as in the non resonant population.

The diffusion in $(e_{m},i_{m})$ generates misidentification of
resonant asteroids belonging to the Vesta and Nysa-Polana families,
whose proper elements differ from the nominal values of the families.
The Massalia dynamical family is less affected by the resonance, probably
due to the low proper inclination of its members that protects them
from large diffusion in eccentricity and inclination. \citet{vokr06}
modeled the orbital evolution of Massalia family members and found
that 16\% of the members that cross the 1:2M resonance are misidentified
due to the effect of the resonance. This is in good agreement with
our result from Table \ref{tabla0}, indicating that inside the resonance
about 11\% of the Massalia family members would not be detected as
such.

When the resonant asteroids reach the chaotic region at $e_{m}>0.22$,
they escape from the resonance mostly by diffusion in $a$. A small
fraction remains captured at higher eccentricities, but because of
their large amplitude of $\sigma$ they undergo encounters with Mars
and escape from the resonance too. Escapes by reaching the secular
resonance $\nu_{6}$ were also observed. For the small asteroids,
the escape by diffusion in $a$ is the rule, but for the big ones
it first operates a diffusion in eccentricity, until they reach $e_{m}>0.22$,
and then operates the diffusion in $a$. A fraction of approximately
6\% of the long lived asteroids end by colliding with the Sun, mainly
after reaching the 3:1 resonance with Jupiter. Asteroids that tend
to follow this path are the ones with the higher inclinations.

\citet{ca10} discussed the origin of the NEA (101955) 1999 RQ36,
with a diameter of 580 m, and concluded that it comes from the Polana
branch of the Nysa-Polana complex, by reaching the $\nu_{6}$ resonance
due to the Yarkovsky effect over long timescales. According to our
results, we suggest another possible path to transfer asteroids from
the Polana region to a NEA orbit: over a small timescale, an asteroid
could reach the 1:2M resonance, excite its eccentricity by diffusion
while evolving in resonant motion, and finally reach the $\nu_{6}$
or the 3:1 resonances on a much shorter timescale than the one proposed
by \citet{ca10}, because for higher eccentricities the diffusion
in semimajor axis is faster, and the secular resonance $\nu_{6}$
is closer to the Polana branch.

\begin{figure}[H]
\resizebox{\hsize}{!}{\includegraphics{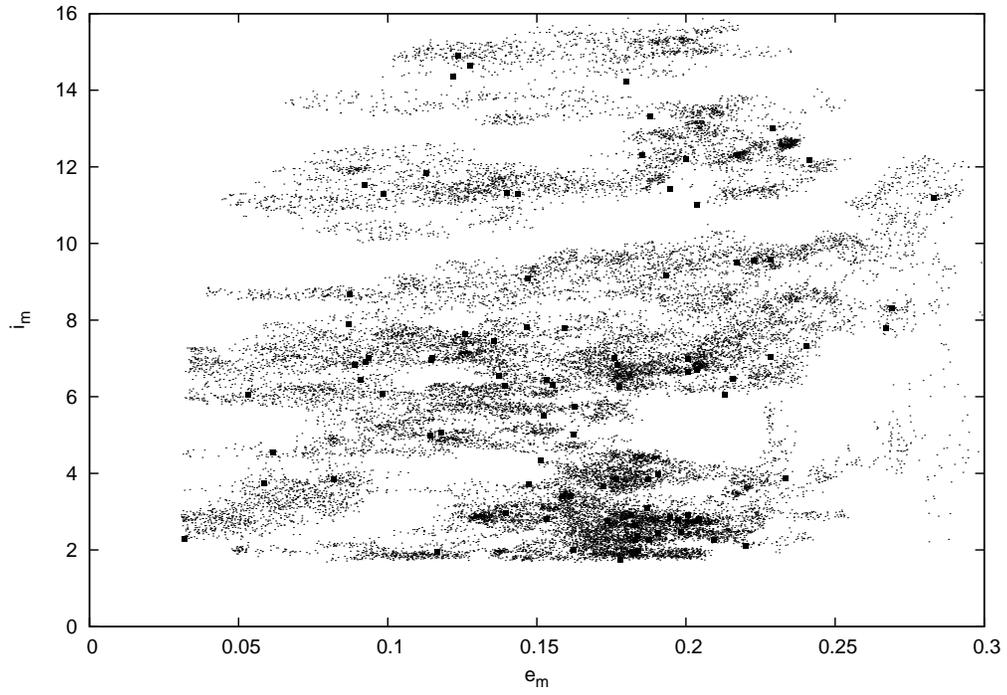}}
\caption{Diffusion in $(e_{m},i_{m})$ of the sample of 100 asteroids shown
in Fig. \ref{limites100}. Diffusion is approximately proportional
to the time span of resonant evolution. Squares indicate the initial
values of the orbits. (Low resolution version)}

\label{einoy}
\end{figure}

\begin{figure}[H]
\resizebox{12cm}{!}{\includegraphics{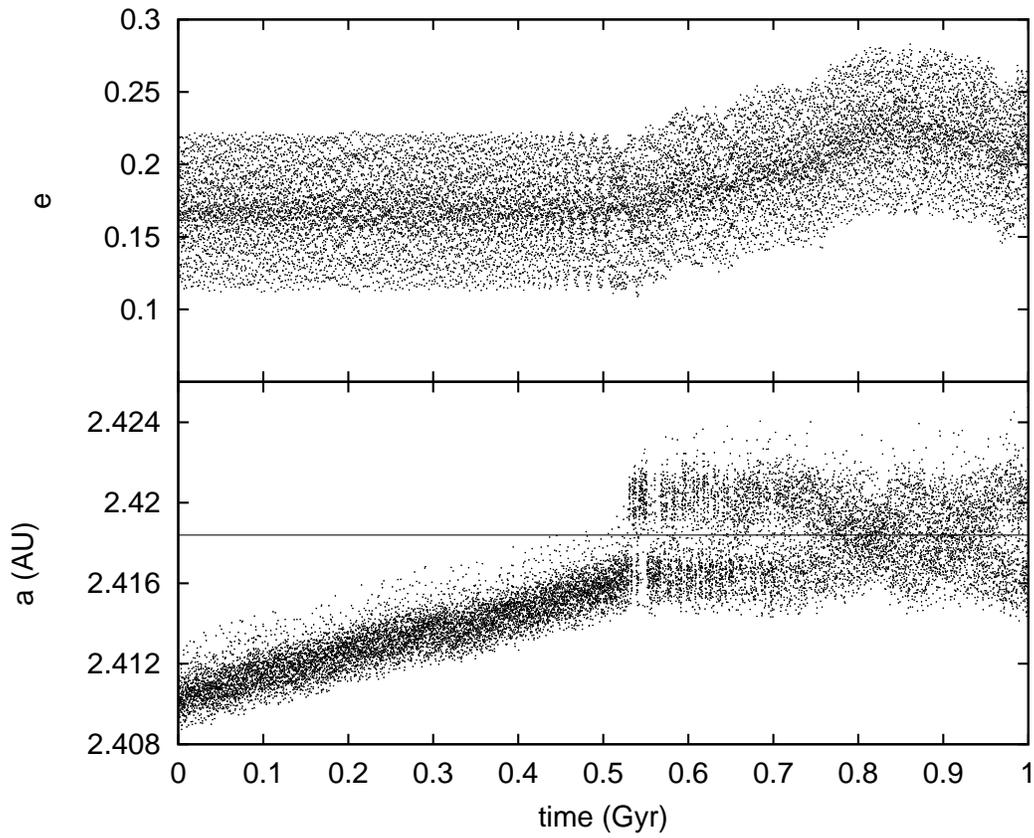}}
\caption{Simulation including a Yarkovsky effect corresponding to a fictitious
asteroid with a radius of some kilometers (simulation \#4). The asteroid
suffers a drift in semimajor axis up to get sticked to the resonance,
then starting a diffusive process in $e$. The exact location of the
resonance is indicated with a line.}

\label{p1119}
\end{figure}

\begin{figure}[H]
\resizebox{\hsize}{!}{\includegraphics{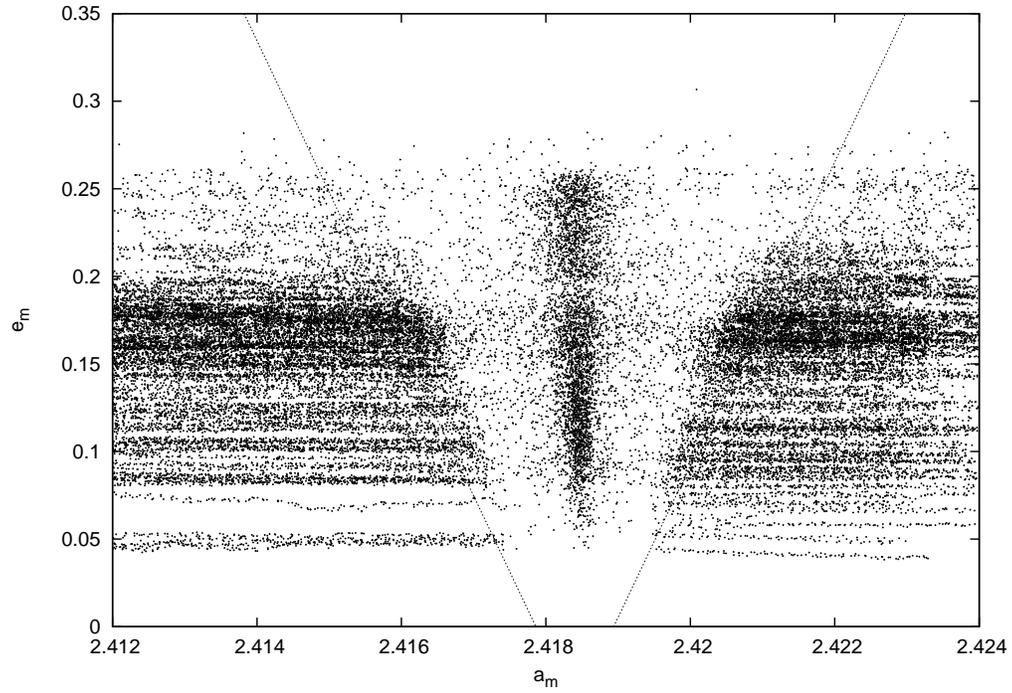}}
\caption{The evolution of the whole population of 200 particles in the simulation
with a Yarkovsky effect corresponding to an asteroid with a radius
of some kilometers (simulation \#4), initially located at the right
side of the resonance and with $\dot{a}<0$. Dotted lines correspond
to the limits of the resonance as computed in Fig. \ref{limites100}. (Low resolution version)}

\label{1gright}
\end{figure}

\begin{figure}[H]
\resizebox{\hsize}{!}{\includegraphics{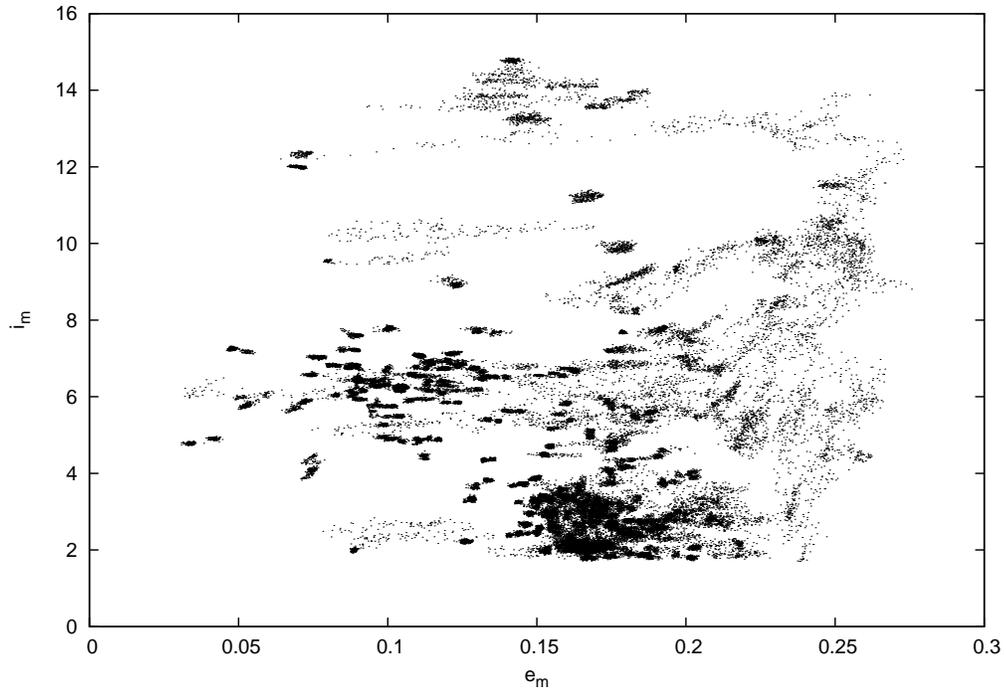}}
\caption{The evolution of the whole population of 200 particles in the simulation
with a Yarkovsky effect corresponding to an asteroid with a radius
of some kilometers (simulation \#4), initially located at the left
side of the resonance and with $\dot{a}>0$. Note the lower diffusion
compared to Fig. \ref{einoy}. Particles that remain captured into
the resonance over a short period of time, or not captured at all,
do not show diffusion in $(e_{m},i_{m})$. Note also the diffusion
in inclination for $e_{m}>0.22$. (Low resolution version)}

\label{1gleft2}
\end{figure}

\begin{figure}[H]
\resizebox{\hsize}{!}{\includegraphics{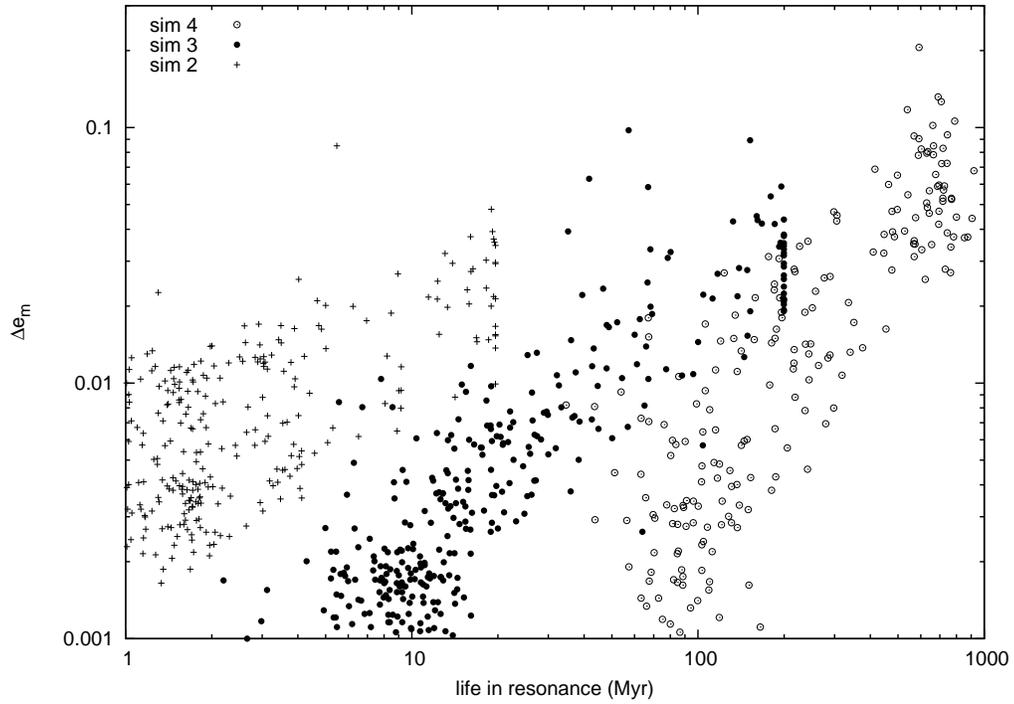}}
\caption{Total variations in $e_{m}$ as a function of the total lifetime in
the resonance, $t_{\mathrm{res}}$, for the simulations \#2, \#3 and
\#4, starting from both the left and right sides of the resonance.
The weaker the Yarkovsky effect, the longer the time evolving in the
resonance and the larger the total diffusion in eccentricity. The
total diffusion in eccentricity follows an approximate power law with
$t_{\mathrm{res}}$.}

\label{sim4der}
\end{figure}

\begin{figure}[H]
\resizebox{\hsize}{!}{\includegraphics{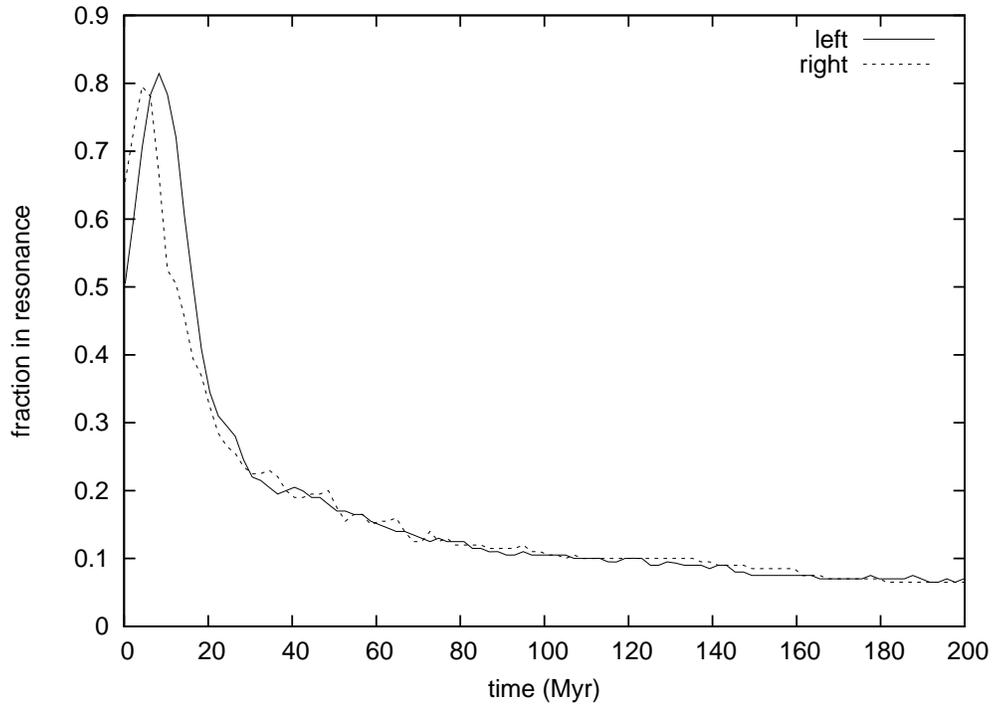}}
\caption{Fraction of the population captured into the resonance, differentiating
between the left and right sides populations, in the case of simulation
\#3 (i.e., with a Yarkovsky effect corresponding to asteroids with
radii of some hundreds meters). All the four simulations (see Table
\ref{tabla1}) show a similar behavior: the evolution from the left
and right populations is indistinguishable, and the tails of long
lived asteroids in both cases are approximately the same.}

\label{evo200}
\end{figure}

The mean lifetime in the resonance of the largest members of the resonant
population, i.e., those not affected by the Yarkovsky effect, is quite
long. These objects can stay evolving in the resonance over timescales
of 1 Gyr, even considering the effects of collisions with other asteroids
and/or ejections due to close encounters with massive asteroids. This
must leave some signature in the orbits of asteroids with diameters
larger than 10 km, because after 1 Gyr of resonant evolution, the
change in eccentricity should be of the order of $\sim0.04-0.15$
(Fig. \ref{einoy}). On the other hand, small resonant asteroids should
have been captured in more recent times.

The negligible effect of the Yarkovsky drift on a large asteroid like
(142) Polana raises the problem of the origin of its present orbit.
There is an analogue case in the resonance 7:3 with Jupiter: asteroid
(677) Aaltje \citep{tsig03}. One possible mechanism to explain such
cases would be chaotic diffusion enhanced by close encounters with
massive asteroids which, over Gyr timescales, might push the big asteroids
into the resonance. A more efficient alternative, although not yet
proved to be compatible with the dynamical history of Mars, would
be a slow migration of Mars orbit. This later mechanism of capture
is dynamically possible, as we have checked with some numerical experiments,
but it does not distinguish between asteroid sizes.

The existence of different $H$ distributions in the resonant and
non resonant populations, with an excess of small asteroids inside
the resonance compared to the background, is an indication that the
main mechanism that replenishes the resonance at present is the Yarkovsky
effect, rather than either chaotic diffusion, close encounters with
massive asteroids, or Mars orbital migration.

\bigskip{}

\textbf{Acknowledgments}

We are grateful to E. Falco for the support provided at the earlier
stages of the numerical integrations, and to M. Cañada by their analysis
of the SDSS data. We acknowledge the criticism and suggestions given
by M. Brož and another anonymous reviewer, that contributed to improve
this work. This study was developed with partial support by PEDECIBA,
and in the framework of the project {}``Caracterización de las Poblaciones
de Cuerpos Menores del Sistema Solar\textquotedbl{} (ANII FCE 2007
318). F.R. acknowledges support by CNPq.

\bigskip{}

\textbf{Appendix: the excess of small asteroids in the resonance}

\begin{figure}[H]
\resizebox{\hsize}{!}{\includegraphics{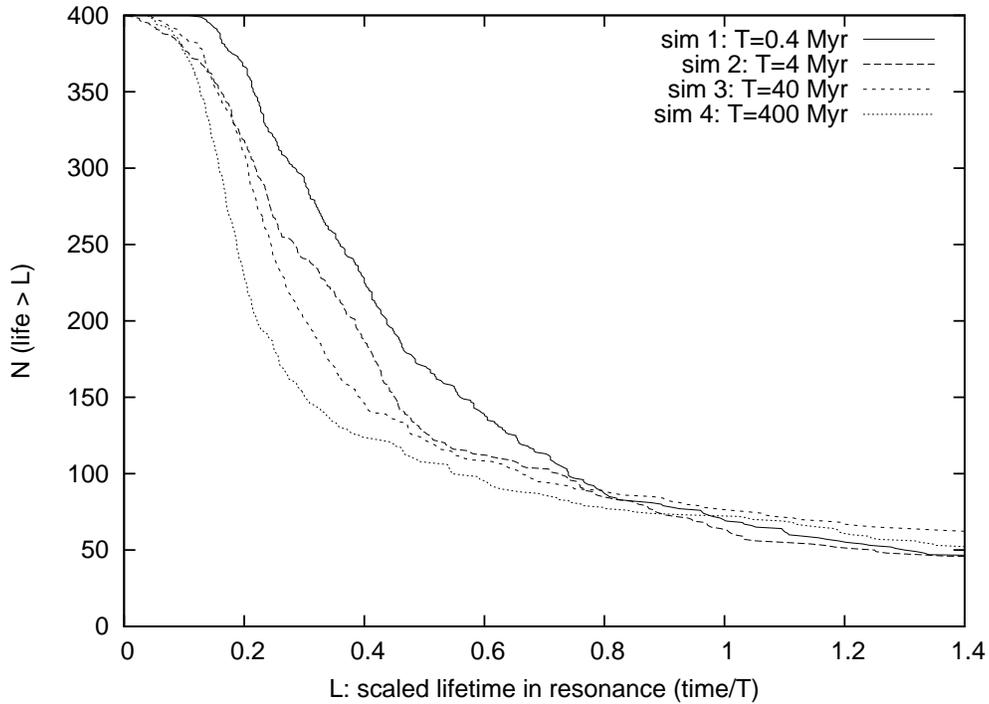}}
\caption{Cumulative distribution of the scaled lifetime in the resonance $L$
for all the simulations with Yarkovsky effect. The left and right
populations appear merged. The lifetime in the resonance is scaled
according to the time $T$ necessary to drift by 0.004 AU, which is
an adopted mean width for the resonance. The decay of the populations
is proportional to the Yarkovsky effect, but in the scaled time the
decay is approximately similar for all the simulations, especially
at the right part of the plot (i.e. long lived asteroids). $L>1$
corresponds to asteroids that have their migration slowed, on average,
by the resonance, whereas $L<1$ corresponds to asteroids that have
their migration accelerated by the resonance.}

\label{4vidas}
\end{figure}

\begin{figure}[H]
\resizebox{\hsize}{!}{\includegraphics{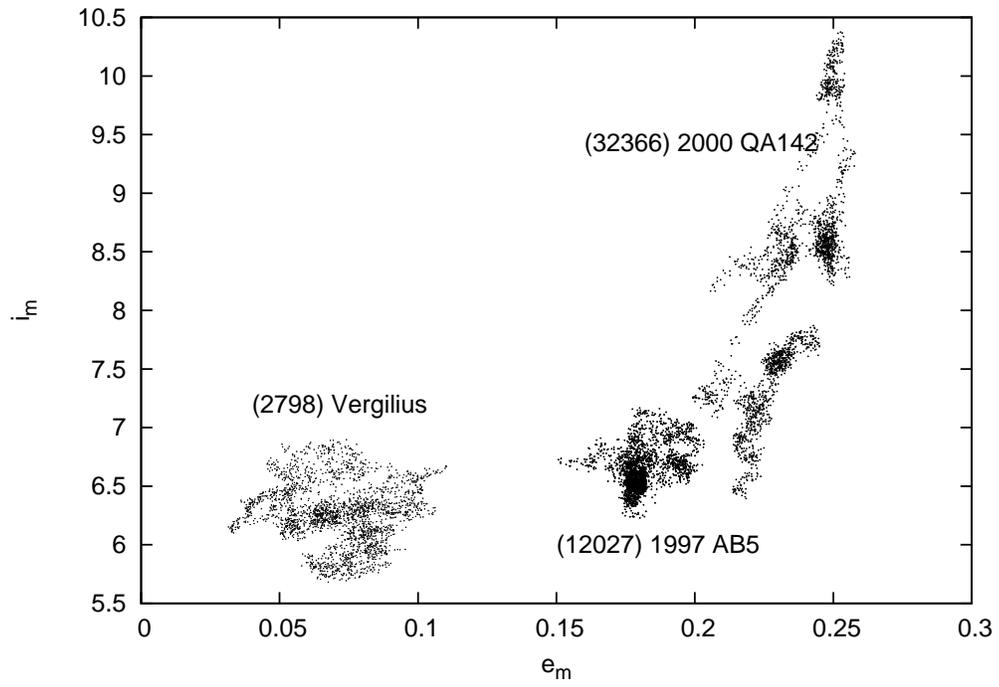}}
\caption{Diffusion over 1 Gyr of three candidate V-type asteroids that evolve
inside the 1:2M resonance and were not identified as members of the
Vesta dynamical family. These asteroids are included in the sample
of 100 asteroids whose evolution is shown in Fig. \ref{einoy}.}

\label{vesto}
\end{figure}

We have defined  $L=t_{\mathrm{res}}/T$, where $T$ is the time that would be necessary to cross the resonance
region due to the Yarkovsky drift in $a$ only.
Taking a mean width for the resonance of $0.004$ AU we have $T=0.004  \hspace{1mm} \textrm{AU} / \dot{a}$.
In steady state, the number of resonant asteroids, $n$, is
\begin{equation}\label{n}
    n=\dot{N}<t_{\mathrm{res}}>
\end{equation}
where
\begin{equation}\label{dotN}
    \dot{N}=\frac{dN}{da}   \hspace{1mm} \dot{a}
\end{equation}
is the injection's rate
of asteroids into the resonance and $<t_{\mathrm{res}}>$ the mean lifetime in resonance,
both  depending on the asteroid size.
By definition of $L$ we have  $<t_{\mathrm{res}}> = <L> T$ for each of the four simulations.
Now, taking into account Eqs. \ref{n} and  \ref{dotN} and the definition of $T$, we can calculate the ratio $n_s/n_l$ between small and large asteroids inside the resonance
\begin{equation}\label{frac}
\frac{n_{s}}{n_{l}}= \frac{dN_{s}}{dN_{l}} \frac{<L_{s}>}{<L_{l}>}
\end{equation}
where $dN_{s}/dN_{l}$ is the ratio outside the resonance.
The behavior of $L$ as seen in Fig. \ref{4vidas} is not very different for the four simulations
at the right part of the plot, corresponding to long lived resonant asteroids.
But, looking at the left part of the plot ($L<1$) it is evident that $L_{s}/L_{l}>1$
then, an excess of small asteroids inside the resonance with respect to outside is expected.

\clearpage{}

\clearpage{}

\clearpage{}

\clearpage{}

\clearpage{}

\clearpage{}

\end{document}